\newcommand{\mywidth}{8cm}

\documentstyle[aps,epsfig,twocolumn]{revtex} 			
\begin{document}

\draft
\title{Reduction of ordered moment and N\'{e}el temperature of quasi 
one-dimensional antiferromagnets Sr$_2$CuO$_3$ and Ca$_2$CuO$_3$}
\author{K.~M.~Kojima$^{1)}$, Y.~Fudamoto$^{1)}$, M.~Larkin$^{1)}$, G.~M.~Luke$^{1)}$, 
J.~Merrin$^{1)}$, B.~Nachumi$^{1)}$, Y.~J.~Uemura$^{1)}$,
N.~Motoyama$^{2)}$, H.~Eisaki$^{2)}$, S.~Uchida$^{2)}$,
K.~Yamada$^{3)}$, Y.~Endoh$^{3)}$, S.~Hosoya$^{4)}$,
B.~J.~Sternlieb$^{5)}$, and G.~Shirane$^{5)}$}

\address{1) Department of Physics, Columbia University, New York, NY 10027, USA \\
2) Department of Applied Physics, University of Tokyo, Tokyo 113, JAPAN\\
3) Department of Physics, Tohoku University, Sendai 980-77, JAPAN\\
4) Institute of Inorganic Synthesis, Yamanashi University, Kofu 400, JAPAN\\
5) Brookhaven National Laboratory, Upton, NY 11973-5000, USA}
\date{submitted on October 10 1996; resubmitted on January 3 1997}
\maketitle
\vspace{0.8cm}
\begin{abstract}
We report elastic neutron diffraction and muon spin relaxation
($\mu$SR) measurements of the quasi one-dimensional antiferromagnets
Sr$_2$CuO$_3$ and Ca$_2$CuO$_3$, which have extraordinarily reduced
$T_{\rm N}/J$ ratios. We observe almost resolution-limited
antiferromagnetic Bragg reflections in Sr$_2$CuO$_3$ and obtain a
reduced ordered moment size of $\sim$0.06$\mu_{\rm B}$.
We find that the ratio of ordered moment size
$\mu($Ca$_2$CuO$_3$$)/\mu($Sr$_2$CuO$_3$$)=1.5(1)$ roughly
scales with their N\'{e}el temperatures, which suggests that the
ordered moment size of quasi one-dimensional antiferromagnets
decreases continuously in the limit of vanishing inter-chain
interactions.
\end{abstract}
\vspace{0.8cm}
\narrowtext

One-dimensional spin systems with antiferromagnetic interactions have
received considerable attention because of their pronounced quantum
mechanical effects. In the absence of inter-chain interactions, both
integer {\it and} half-odd integer spin-chain systems have spin-singlet
ground states, rather than an antiferromagnetically ordered N\'{e}el
state \cite{BetheZPhys31,MarminPRL66,HaldanePL83PRL83}. Yet, for half 
odd-integer spin-chains, the spin-excitations are gap-less
at momentum $k=0\ {\rm and}\ \pi$ \cite{desCloizeauxPR62}; this
indicates that the ground state of a half-odd integer spin-chain is closer 
to the N\'{e}el ordered state than the integer spin systems, which 
have a so-called Haldane gap \cite{HaldanePL83PRL83}.

Because of the gap-less feature of half-odd integer spin-chains,
one interesting question is whether the ground state is ordered or
disordered when inter-chain interactions are introduced.
Previously, it was proposed that there is a non-zero critical
coupling ratio ($J'/J=R_{\rm c}$), below which the system
retains a singlet ground-state \cite{ParolaPRL93}.  Recent
renormalization group calculations however suggest that the ground
state may depend on microscopic details of the model which describes
the spin-spin interactions \cite{AffleckPre96,AffleckJPhys94}.
Numerical studies of the Heisenberg model suggested a vanishing
critical coupling ratio ($R_{\rm c}\sim 0$); namely, for
infinitesimally small inter-chain couplings, half odd-integer
spin-chains should exhibit N\'{e}el  order \cite{AffleckJPhys94}.

Experimentally, KCuF$_3$ is the most investigated
quasi-one-dimensional $S$=1/2 antiferromagnet. Unfortunately, this
material has relatively large coupling ratio $R=J'/J\sim 2\,{\rm
K}/203\,{\rm K}=1.0\times 10^{-2}$, as shown from neutron
inelastic scattering measurements \cite{SatijaPRB80}. Probably reflecting 
the large coupling ratio $R$, 
the $T_{\rm N}/J$ ratio ($\sim 39\, {\rm K}/203\, {\rm K}= 0.2$) 
and the ordered moment size ($=0.49(7) \mu_{\rm B}$ \cite{HutchingsPR69})
were also found to be relatively large.

To investigate the regime of the critical coupling ratio, model
materials with smaller inter-chain couplings are needed; the quasi
one-dimensional $S$=1/2 antiferromagnets Sr$_2$CuO$_3$ and
Ca$_2$CuO$_3$ are suitable candidates. The intra-chain interaction
($2J\sim 2600$~K) of these materials have been estimated from susceptibility
\cite{AmiPRB95,MotoyamaPRL96} and infrared light absorption \cite{SuzuuraPRL96}.
N\'{e}el ordering of these compounds was first observed in $\mu$SR\ measurements
\cite{KerenPRB93JMMM95a}, with a significantly reduced $T_{\rm
N}/J$ ratio of $\sim 5\,{\rm K}/1300\, {\rm K}= 4\times 10^{-3}$ for
Sr$_2$CuO$_3$ and $T_{\rm N}/J\sim 11\, {\rm K}/1300\, {\rm K}=
8\times 10^{-3}$ for Ca$_2$CuO$_3$.  Since $T_{\rm N}$/$J$ is a measure of
the coupling ratio $R$ \cite{AffleckJPhys94,OguchiPR64}, the reduced
$T_{\rm N}$\ of these two compounds demonstrates their good
one-dimensionality.  A previous elastic neutron scattering measurement
of Ca$_2$CuO$_3$ \cite{YamadaPHYSCA95} has found an extremely reduced
size of ordered moments (=0.05(3)$\mu_{\rm B}$), although this result
contains a systematic uncertainty due to extinction.  In the case of
Sr$_2$CuO$_3$, powder neutron measurements were unable to observe
antiferromagnetic Bragg reflections, placing an upper limit of any
ordered moment of $<0.1 \mu_{\rm B}$ \cite{AmiPRB95}. In this Letter
we report $\mu$SR and neutron scattering measurements of single
crystalline Sr$_2$CuO$_3$ and Ca$_2$CuO$_3$ specimens, aiming to
clarify the relationship between $T_{\rm N}/J$ and the size of ordered
moments.

\begin{figure}[h,t]
\begin{center}
\mbox{\epsfig{file=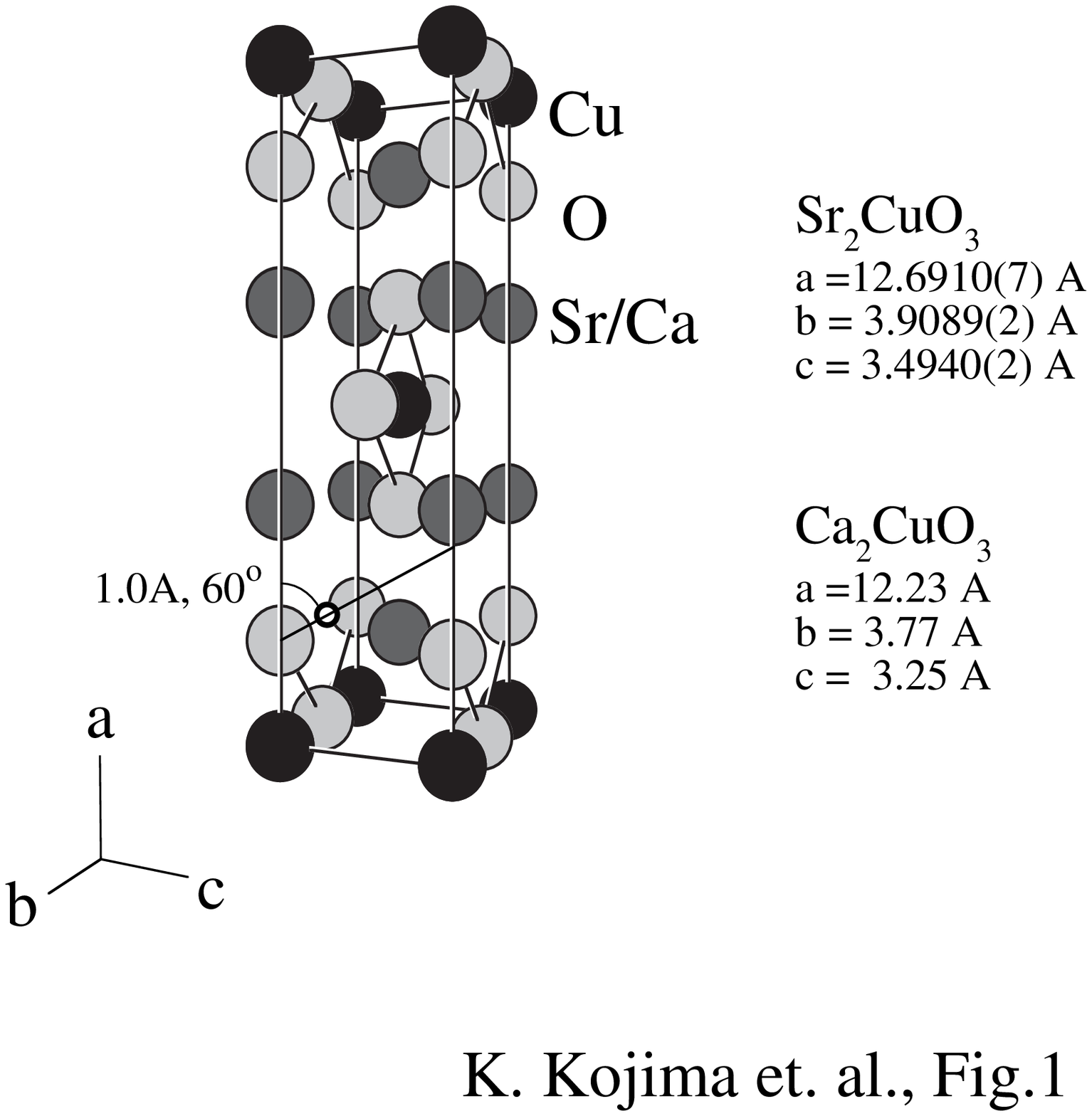,width=\mywidth}}
\end{center}
\caption{}
\small
The crystal structure of (Sr,Ca)$_2$CuO$_3$. The Cu-O chain runs in
the $b$-axis direction. The circle is the off-chain O-$\mu^{+}$ bond
muon site.  The lattice parameters are from
Ref.\cite{AmiPRB95,TeskeZAAC70}.
\normalsize
\label{fig:crystal}
\end{figure}

The crystal structure of Sr$_2$CuO$_3$ and Ca$_2$CuO$_3$\
(Fig.\ref{fig:crystal}) is similar to that of La$_2$CuO$_4$, but
lacks oxygen ions between the Cu ions in one direction ($c$-axis).  As
a result, chains of corner shared CuO$_4$ tetragons extend in the
$b$-axis direction, with a strong antiferromagnetic interaction
due to the 180$^\circ$ Cu-O-Cu coupling. The lattice parameters of
Ca$_2$CuO$_3$ are smaller than those of Sr$_2$CuO$_3$ by 7.0\%
($c$-axis) and 3.6\% ($a$- and $b$-axis)
\cite{AmiPRB95,TeskeZAAC70}.  The reduced $c$-axis
parameter of Ca$_2$CuO$_3$ probably enhances the inter-chain coupling
($J'$), as its higher $T_{\rm N}$\ suggests.

A single crystal of Sr$_2$CuO$_3$ ($\sim\phi 3{\rm mm}\times 2{\rm
cm}$) was grown employing the traveling-solvent-floating-zone (TSFZ)
method, as described in Ref.\cite{MotoyamaPRL96}. In order to search
for antiferromagnetic Bragg reflections, we performed elastic neutron
scattering measurements at the High Flux Beam Reactor (HFBR) at
Brookhaven National Laboratory, using the H4M and H7 triple-axis
spectrometers. For the measurements, two pyrolytic graphite (PG) 
filters were employed to eliminate contamination of higher order
reflections from the monochromator.

\begin{figure}[h,t]
\begin{center}
\mbox{\epsfig{file=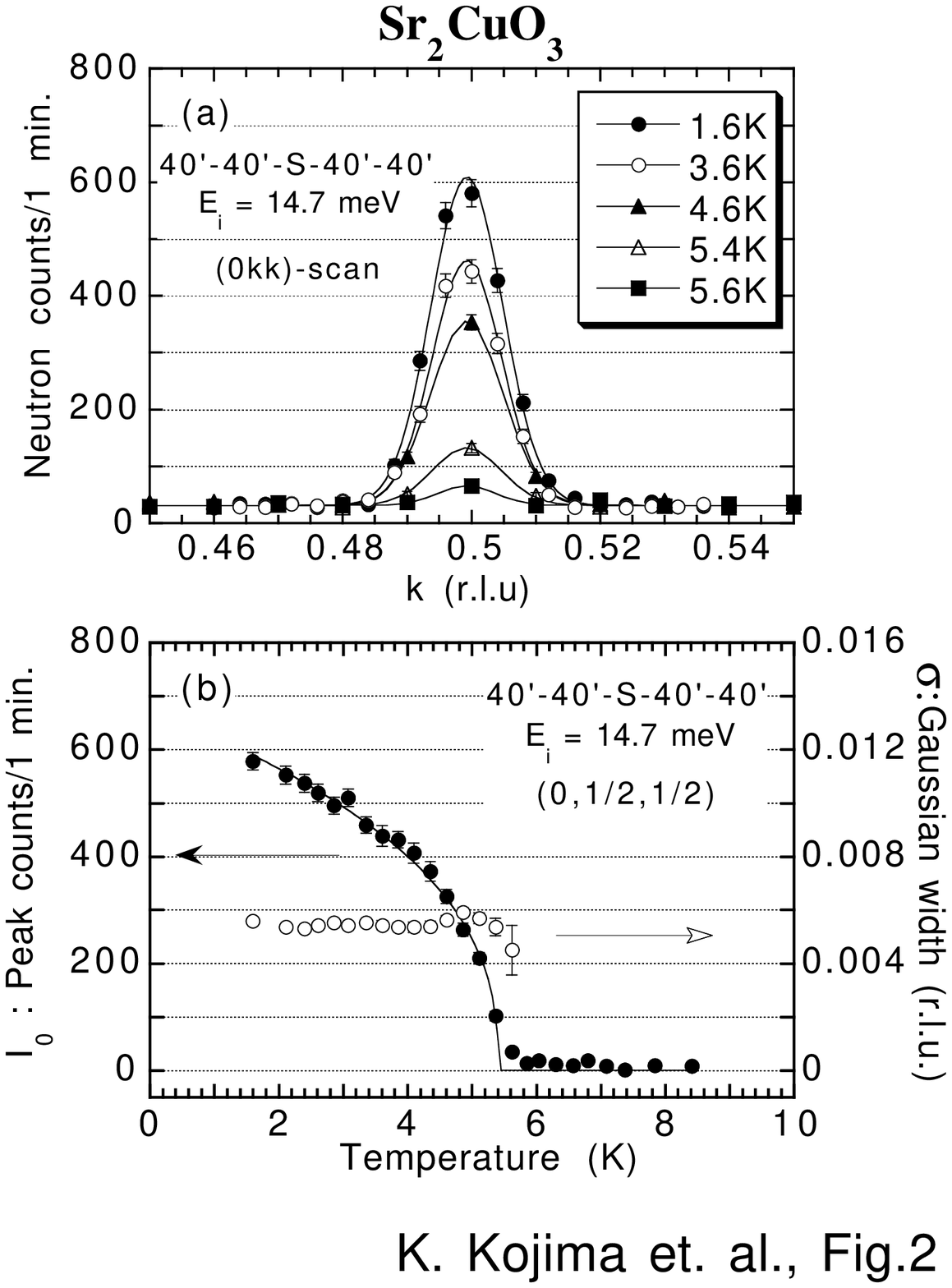,width=\mywidth}}
\end{center}
\caption{}
\small
(a) Antiferromagnetic Bragg reflection of Sr$_2$CuO$_3$. (b)
Temperature dependence of the peak counts $I_0$ (filled circles), and
the width $\sigma$ (open circles).  The solid line is a
phenomenological power-law fit ($I_0\propto (T_{\rm N}-T)^{2\beta}$)
with $T_{\rm N}= 5.41(1)$~K and $\beta=0.20(1)$.
\normalsize
\label{fig:AFBragg}
\end{figure}

In Fig.\ref{fig:AFBragg}a, we show diffracted neutron counts around
the point $(0,1/2,1/2)$, where an antiferromagnetic Bragg reflection
was observed below $T_{\rm N}=5.41(1)$K. We confirmed with tighter
collimation (10'-40'-S-10'-80') that the width of this Bragg
reflection is as narrow as that of a nuclear reflection (011).  This
is direct evidence of antiferromagnetic long range order in
Sr$_2$CuO$_3$.  We observed other antiferromagnetic Bragg reflections
at $(h,k/2,l/2)$, where $h$ is an integer and $k\ {\rm and}\ l$ are
odd-integers. We fit the $(0,1/2,1/2)$ reflection with a Gaussian
form, and plot the peak intensity ($I_0$) and width ($\sigma$) in
Fig.\ref{fig:AFBragg}b as a function of temperature.  

In order to determine the ordered spin-direction and the moment size,
we measured the integrated intensities of magnetic Bragg reflections
in both the $0kl-$ and $hkk-$zones.  The intensity distribution was
best explained using the assumption that ordered moments are aligned
along the $b$-axis direction, parallel to the chain.  By
normalizing the magnetic Bragg intensities with those of several
relatively weak nuclear Bragg reflections, such as (200) and (400), we
find the ordered moment size to be 0.06(3) $\mu_{\rm B}$.  Because of
extinction of nuclear reflections, the ordered moment size obtained here should
be considered as an upper limit.

Since extinction of nuclear reflections depends on the quality of
individual crystals, the size of ordered moments obtained by the above
method contains a relatively large systematic error. With muon spin
relaxation ($\mu$SR) on the other hand, we can compare the relative size of
moments between iso-structural materials quite accurately.  Muons in
Sr$_2$CuO$_3$ and Ca$_2$CuO$_3$ are expected to occupy the same
crystallographic position and experience dipolar fields from the ordered
moments below $T_{\rm N}$.  The relative size of ordered moments can
be deduced from the muon spin precession frequencies.

The $\mu$SR measurements were performed at the M15 surface-muon
channel at TRIUMF (Vancouver, Canada), using a conventional $\mu$SR
spectrometer \cite{HayanoPRB79} combined with a dilution refrigerator
and a `low-background' apparatus \cite{ArseneauHI96} with a $^4$He
gas-flow cryostat.  We evaluated the time evolution of muon spins,
using the conventional ZF/LF-$\mu$SR technique
\cite{HayanoPRB79,UemuraPRB85}.

\begin{figure}[h,t]
\begin{center}
\mbox{\epsfig{file=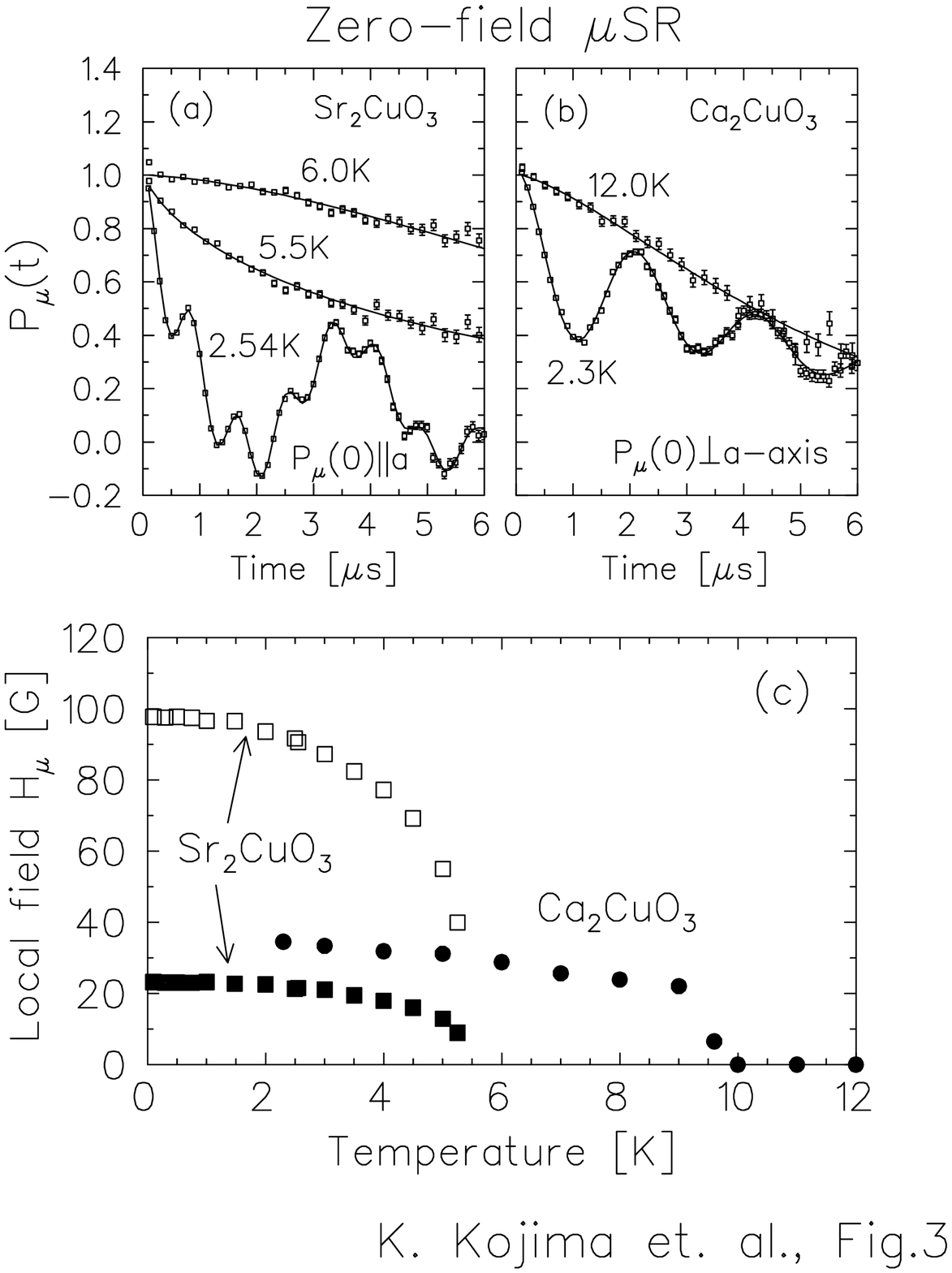,width=\mywidth}}
\end{center}
\caption{}
\small
Zero-field $\mu$SR spectra of (a) Sr$_2$CuO$_3$ and (b) Ca$_2$CuO$_3$.
The solid lines on the data below $T_{\rm N}$ are fits to the function
described in the text.  (c) The local fields at the muon sites
($H_{\mu 1,2}$) are shown.
\normalsize
\label{fig:muSRA2CuO3}
\end{figure}

In Fig.\ref{fig:muSRA2CuO3}a, we show the spectrum for Sr$_2$CuO$_3$.
Below $T_{\rm N}$, we observe spontaneous muon spin precession in
zero external magnetic-field; this is a signature of a well-defined
static local field from ordered moments.  We analyzed the spectra
assuming two muon sites:
\begin{eqnarray}
P_\mu(t)&=&A_1P_{\mu 1}(t)+A_2P_{\mu 2}(t)
\label{eq:Pmu}
\end{eqnarray}
where $A_{1,2}$ is the fractional site population of the muons ($A_1+A_2=1$). 
We assumed the following conventional form for the signal from each site:
\begin{eqnarray}
P_{\mu i}(t) &=&A_{{\rm osc}i}\exp(-\Delta_i t)\cos(\gamma_\mu H_{\mu i}t+\phi_i)\nonumber\\
           & &+A_{{\rm rlx}i}\exp(-t/T_{1 i})\;\;\;\;\;\;(i=1,2)
\label{eq:Pmui}
\end{eqnarray}
where, the first (second) term presents the muon spin precession
($T_1$ relaxation), due to the local field component perpendicular
(parallel) to the initial muon spin direction. 
In Fig.\ref{fig:muSRA2CuO3}c, we show the local fields ($H_{\mu 1,2}$)
as a function of temperature. In Sr$_2$CuO$_3$, the ratio of the two
local fields was independent of temperature; this suggests that both
of the muon sites are stable and that muons do not hop on the
time scale of the muon lifetime.  In Fig.\ref{fig:muSRA2CuO3}b, we
show $\mu$SR spectra of Ca$_2$CuO$_3$.  Because of the shape of our
Ca$_2$CuO$_3$ specimen, we performed the $\mu$SR measurements with a
crystal orientation ($P_{\mu}(0) \perp a$-axis) which was different
from that of the Sr$_2$CuO$_3$ case ($P_\mu(0)\parallel\ a$-axis
$\perp$ chain).  Consequently, we observed only one signal in the
ordered phase. We confirmed, from independent measurements of
polycrystalline Ca$_2$CuO$_3$ pellets, that the higher frequency
signal also exists and that the signal observed in the single crystalline
sample corresponds to the lower frequency signal. The muon local
fields are plotted in Fig.\ref{fig:muSRA2CuO3}c.  The ratio of the
local fields of the two systems, which is equal to the relative size
of ordered moments, was
$\mu($Ca$_2$CuO$_3$$)/\mu($Sr$_2$CuO$_3$$)=35(3) {\rm G}/23.2(1) {\rm
G}=1.5(1)$ in the $T\rightarrow 0$ limit.

In high-$T_{\rm c}$ related oxides, muons generally form an
O-$\mu^{+}$ bond with a bond length of $\sim 1.0$~\AA
\cite{muonsites}.  Assuming such O-$\mu^{+}$ bond formation,
we performed an electrostatic potential calculation, and determined
the stable muon positions in (Sr,Ca)$_2$CuO$_3$. Fig.\ref{fig:crystal}
shows the off-chain O-$\mu^{+}$ bond site, which is responsible for
the lower-field signal. We calculated the magnetic dipolar-field for
this site, and found that the local field from a given size of ordered
moment agrees within 10\% in Sr$_2$CuO$_3$ and Ca$_2$CuO$_3$.
Therefore, the local field ratio in these two compounds reflects the
relative size of their moments. We estimated the ordered moment sizes
from the dipolar-fields, as summarized in Table
\ref{table:localfield}. The moment sizes obtained by $\mu$SR and
neutron techniques agreed within the errors, suggesting that
uncertainties due to extinction or muon site ambiguity are, in fact,
rather small.

\begin{figure}[h,t]
\begin{center}
\mbox{\epsfig{file=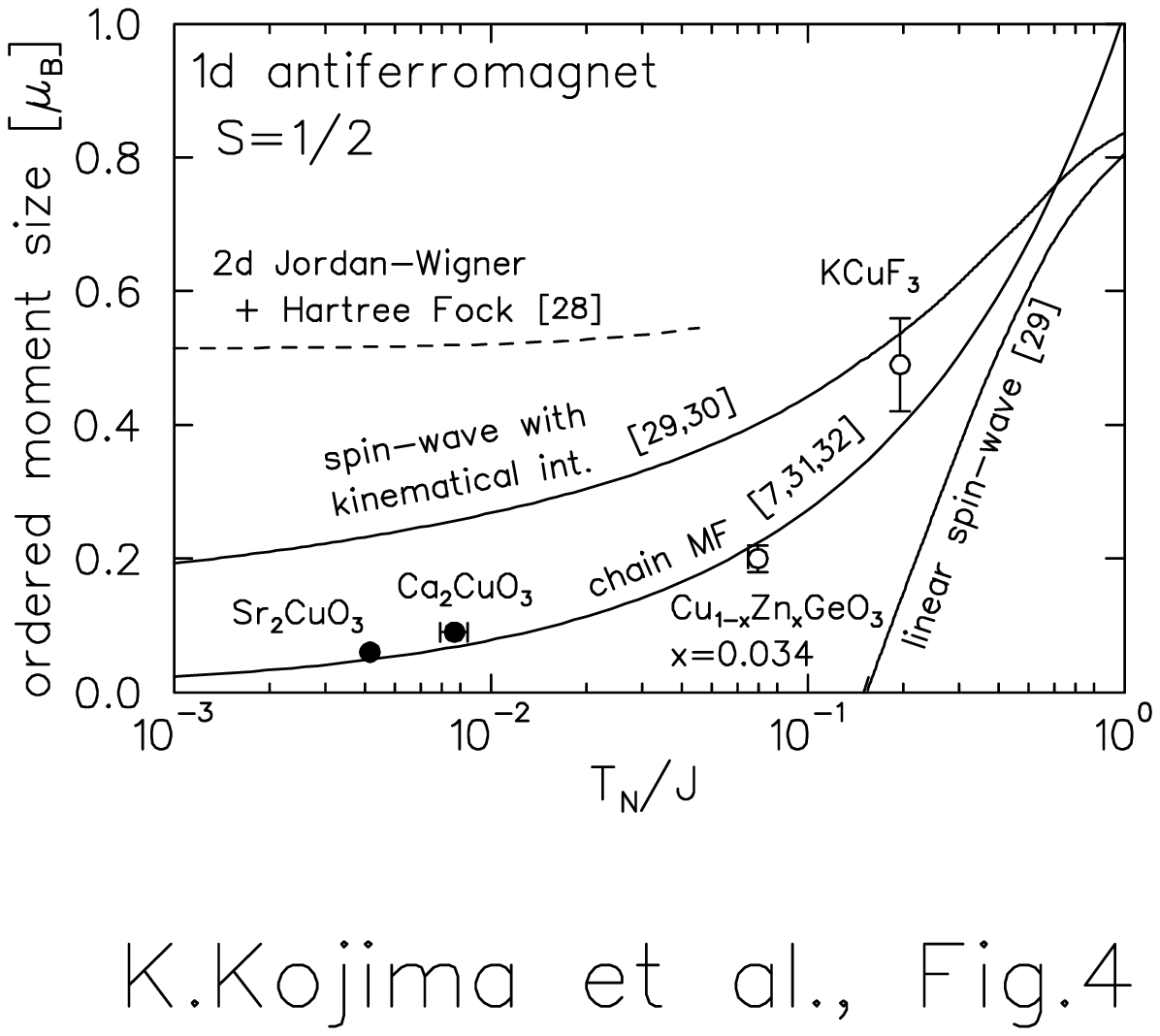,width=\mywidth}}
\end{center}
\caption{}
\small
Ordered moment size as a function of $T_{\rm N}/J$. The points for
Sr$_2$CuO$_3$ and Ca$_2$CuO$_3$ are from this work (see Table
\ref{table:localfield}). KCuF$_3$ is from Ref.
\cite{SatijaPRB80,HutchingsPR69} and Cu$_{1-x}$Zn$_x$GeO$_3$ ($x=0.034$)
is from Ref.\cite{HaseJPSJ96,NishiPRB94}.  The lines are theoretical
relations, which have zero moment (solid lines from Ref. \cite{WelzJPhys93,SchulzPRL96}) 
or finite moment (dashed line from Ref. \cite{AzzouzPRB93}) in the $R\rightarrow 0$ limit. 
\normalsize
\label{fig:Tn-moment}
\end{figure}

In Fig.\ref{fig:Tn-moment}, we plot the ordered moment size of several 
quasi 1d antiferromagnets as a function of $T_{\rm N}/J$. As expected, the
ordered moments shrink as the $T_{\rm N}/J$ ratio decreases. Moreover,
the ordered moment size continuously decreases in the regime of
extremely reduced ratio $T_{\rm N}/J=4\times 10^{-3}\sim 10^{-2}$.
This suggests that the ordered moment vanishes smoothly in the
$R=J'/J\rightarrow 0$ limit, rather than maintaining a limiting size
(dashed line in Fig.\ref{fig:Tn-moment}) as has been proposed
theoretically \cite{AzzouzPRB93}.

The solid lines in Fig.\ref{fig:Tn-moment} are predictions of ordered
moment size from (1) linear spin-wave theory \cite{WelzJPhys93} (2)
spin-wave theory with kinematical interactions
\cite{WelzJPhys93,IshikawaPTPhys75} and (3) chain mean-field (CMF) theory 
\cite{AffleckJPhys94,SchulzPRL96,ScalapinoPRB75}. These theories predict the relationship
between the coupling ratio ($R=J'/J$) and the ordered moment size; we
estimated $R$ from $T_{\rm N}/J$ within the framework of each theory
($T_{\rm N}/J\approx 2.1 S(S+1)\sqrt{J'/J}$ \cite{OguchiPR64} for the
spin-wave theories and $T_{\rm N}/J\approx J'/J$
\cite{AffleckPre96,AffleckJPhys94,SchulzPRL96} for chain mean-field theory.)  We
found that the CMF approach best explains the ordered moment size of
Sr$_2$CuO$_3$ and Ca$_2$CuO$_3$. Since the CMF approach is based on
the exact solution of an isolated spin-chain, it takes more quantum
mechanical effects into account than the spin-wave approaches which
presume an antiferromagnetic ordered state.  Probably, this is the
reason why the CMF approach presented the most successful account for
$R\rightarrow 0$ region, where moment reduction is dominated by
quantum spin fluctuations.

Summarizing our results, we observed long range antiferromagnetic
order in Sr$_2$CuO$_3$ with a remarkably reduced ordered moment ($\sim
0.06 \mu_{\rm B}$).  The relative size of ordered moments
$\mu($Ca$_2$CuO$_3$$)/\mu($Sr$_2$CuO$_3$$)=1.5(1)$ roughly scales with
their N\'{e}el temperatures. We have shown that Sr$_2$CuO$_3$ and
Ca$_2$CuO$_3$ lie in the regime of extremely reduced $T_{\rm N}/J$
ratio, where quantum mechanical moment reduction dominates. Further
studies of these systems, for example, probing their spin excitations
will be of great interest.

We wish to thank Profs. I.~Affleck, H.~Fukuyama, M.~Gingras and
H.~J.~Schulz for very helpful discussions.  We also thank the TRIUMF
$\mu$SR facility for its hospitality. One of the authors (K.K.) is
supported by an JSPS overseas fellowship. This research has been
financially supported by NEDO (Japan) and NSF (DMR-95-10454). The work
at BNL was supported in part by the U.S. Japan cooperation Research
Program on neutron scattering, and by the Division of Material Science
at the U.S. Department of Energy, under contract No.
DE-AC-2-76CH00016.

\begin{table}[h]
\caption{Ordered moment size of 1d and 2d antiferromagnets}
\begin{tabular}{c c c c}
compound & \multicolumn{2}{c}{$\mu$SR} & neutron \\
         & local fields &{moment size}& moment size \\
\hline
Sr$_2$CuO$_3$$^{\rm a}$& 23.2~G, 97.7~G & 0.06(1) $\mu_{\rm B}$ & 0.06(3) $\mu_{\rm B}$\\
Ca$_2$CuO$_3$$^{\rm b}$& 35~G         & 0.09(1) $\mu_{\rm B}$ & 0.05(3) $\mu_{\rm B}$\\
YBa$_2$Cu$_3$O$_7$$^{\rm c}$& 310~G, 1330~G &       & 0.6 $\mu_{\rm B}$\\
La$_2$CuO$_4$$^{\rm d}$     & 430~G       &       & 0.5 $\mu_{\rm B}$\\
\end{tabular}
\vspace{8pt}
\tablenotetext[1]{this work.}
\tablenotetext[2]{$\mu$SR: this work, and neutron: Ref. \cite{YamadaPHYSCA95}.}
\tablenotetext[3]{$\mu$SR: from Ref.\cite{BrewerPHYSCA89}, 
	and neutron: Ref. \cite{TranquadaPRB88}.}
\tablenotetext[4]{$\mu$SR: from Ref.\cite{UemuraPRL87}, 
	and neutron: Ref. \cite{VakninPRL87}.}
\label{table:localfield}
\end{table}

\end{document}